\documentclass[twocolumn,showpacs,preprintnumbers,amsmath,amssymb]{revtex4}
\usepackage{graphics}
\begin{document}
\title{Hydrodynamics of polar liquid crystals}
\author{William Kung}\email{wkung@physics.syr.edu}
\affiliation{Department of Physics, Syracuse University, Syracuse,
New York 13244}
\author{Karl Saunders}\affiliation{Department of Physics, California Polytechnic State University, San Luis Obispo, California 93407}
\author{M. Cristina Marchetti}
\affiliation{Department of Physics, Syracuse University, Syracuse, New York 13244}
\date{\today}

\begin{abstract}
Starting from a microscopic definition of an alignment vector
proportional to the polarization, we  discuss the hydrodynamics of
polar liquid crystals with local $C_{\infty v}$-symmetry.  The
free energy for polar liquid crystals differs from that of nematic
liquid crystals ($D_{\infty h}$) in that it contains terms
violating the ${\bf{n}}\rightarrow -{\bf{n}}$ symmetry. First we
show that these
 $\mathcal{Z}_2$-odd terms induce a general splay
instability of a uniform polarized state in a range of parameters.
Next we use the general Poisson-bracket formalism to derive the
hydrodynamic equations of the system in the polarized state. The
structure of the linear hydrodynamic modes confirms the existence
of the splay instability.
\end{abstract}
\maketitle

\section{Introduction}
In the past few decades, our understanding of complex materials
has greatly benefited from the increasing sophistication of
technology and experimental designs.  Depending upon the various
properties of the constituent molecules, such as geometric shape,
electric and magnetic moments, as well as chemical affinity, we
are able to observe a variety of phases beyond the traditional
crystalline and liquid phases. Examples range from the traditional
nematic and smectic liquid crystal phases~\cite{LubenskyText} to
the recently discovered smectic blue phase~\cite{blue1} or the
banana phases~\cite{Banana1}. The possibility of ferroelectric
order in liquid crystals has attracted much attention both
from a fundamental and practical viewpoint. Among the commonly
observed liquid crystalline phases, only the chiral smectic-C*
phase is known to be polar. Ferroelectricity in this phase was
discovered in 1975~\cite{Meyer} and the related phenomenon of
antiferroelectricity in liquid crystals was discovered in
1989~\cite{Chandani}. In these chiral liquid crystals, polarity is
realized via the introduction of chiral carbon atoms into achiral
molecules.  Ferroelectric switching was also recently confirmed in
the banana smectic phases~\cite{Banana1} consisting of achiral
bend-core molecules.  The packing structure in these banana phases
induces a polar order in the layers of these liquid-crystal
systems.

The possibility of ferroelectric and antiferroelectric order in
achiral polar liquid crystals composed of molecules carrying
strong permanent electrical dipole moments has been a
long-standing theoretical and experimental question.
Experimentally, ferroelectric order due to dipole-dipole
interactions between rod-like aromatic copolyesters molecules has
been reported~\cite{exp1,exp2,exp3,exp4,exp5} by the Tokyo Tech
group. In this case the nematic liquid crystal is biaxial and
polarity was observed along both symmetry axes via second-harmonic
generation.  Analytical (generally mean-field) and numerical
calculations~\cite{Paffy, Mettout, Lee, Alexander,Baus,Teixeira1,
Park}, as well as simulations~\cite{Wei, Teixeira2, Biscarini,
Weiss}, have argued that there is no fundamental reason forbidding
the establishment of ferroelectric order in polar liquid crystals,
assuming appropriate surface stabilization.

The simplest ferroelectric liquid crystal of this type would be a
uniaxial liquid of molecules carrying a permanent electric dipole
moment which lies parallel to the molecular axis~\cite{banana}. In
this case ferroelectricity would be due solely to dipolar
interactions, with all dipoles pointing in the same direction in
the ordered state. In their classic textbook de Gennes and Prost
argue that such dipolar molecules would always be asymmetric in
shape, with head and tails of different size. This would naturally
yield a splay deformation, making the liquid ferroelectric state
unstable~\cite{DeGennestext}. Theoretical work has indeed shown
that any type of molecular asymmetry will suppress ferroelectric
order in nematic liquid crystals, possibly replacing it with
modulated polar phases~\cite{Paffy}. In this paper we show that
the instability of an ordered ferroelectric phase of uniaxial
polar liquid crystals is  more general than originally suggested
by De Gennes and Prost.  We demonstrate that in uniaxial polar
liquid crystals a ferroelectric phase is theoretically possible
but only for a limited range of parameters. Outside this range the
uniformly polarized state is unstable to splay distortions,
regardless of molecular shape. This may explain the difficulties
encountered in establishing experimentally the existence of a
ferroelectric polar state. We employ a standard Poisson bracket
formalism~\cite{Volovik80, Volovik80b} to derive the hydrodynamic equations for a liquid of
uniaxial rod-like molecules carrying permanent dipole moments.
From these equations we obtain the hydrodynamic modes in the
uniformly polarized state and show that in a range of parameters
the ordered state becomes unstable against the growth of
long-wavelength splay fluctuations. This is consistent with an
analysis of the free energy which indicates the presence of a
mechanical instability for the same parameter range.  Our
derivation is based solely on symmetry arguments and makes no
assumption on the shape of the molecules.

The familiar nematic liquid crystal phase is characterized by a
broken rotational symmetry due to macroscopic order of the
rod-like molecules along a fixed direction in space. The broken
symmetry direction is identified by a unit vector ${\bf{n}}$,
known as the director. The states of director ${\bf{n}}$ and
$-{\bf{n}}$ are indistinguishable. If the molecules carry a
permanent electric dipole moment, in the ordered state there as
many 'up' dipoles as 'down' dipoles and the system is not
ferroelectric on large scales. The main difference between a polar
liquid crystal and the familiar nematic phase is the breaking of
such a ${\mathcal{Z}}_2$-symmetry $({\bf{n}}\rightarrow
-{\bf{n}})$. If the molecules carry a dipole moment, on average
the dipoles are pointing in the same direction in the ordered state
and the system can be ferroelectric on macroscopic scales. The order
parameter characterizing the transition from the isotropic liquid
phase to the polarized phase is the alignment vector
${\sl{\bf{P}}}$, proportional to the mean polarization of the
system. This should be contrasted to the nematic order parameter
$Q_{ij}$, which is a rank-two tensor, also known as the alignment
tensor. In uniaxial systems the mean values of both the alignment
vector and the alignment tensor in an ordered phase can be
expressed in terms of a unit vector pointing along the broken
symmetry direction,
\begin{eqnarray}
&&{\bf P} =P_0{\bf n}\;,\\
&&Q_{ij}=S\Big(n'_in'_j-\frac{1}{3}\delta_{ij}\Big)\;.
\end{eqnarray}
The unit vectors ${\bf n}$ and ${\bf
n'}$ are the directors in the polarized and nematic states,
respectively. A macroscopic polarized state is characterized by
$P_0\not=0$. From this it follows that $S\not=0$ as well in the
polarized state, and ${\bf n'}={\bf n}$. The nematic phase has
$P_0=0$, but $S\not=0$. In the following we derive hydrodynamic
equations for a polar liquid crystal. The hydrodynamic fields in
the isotropic phase are simply the five conserved densities
(number density, momentum density and energy density). In the
polarized phase we must add the two broken symmetry variables
corresponding to the two independent components of the director
${\bf n}$. In addition, we obtain the dynamical equation for the full
alignment vector ${\bf P}$, which determines the order parameter
for the polarized-isotropic phase transition. Its magnitude is not a
hydrodynamic variable, but it plays an important role in
describing the dynamics near the transition, where its relaxation
rate may become very slow.

Our derivation starts in section II with the microscopic
definitions of the relevant hydrodynamic fields. Using standard
definitions from classical mechanics, in section III we obtain the
coarse-grained Poisson-brackets for these macroscopic fields.
Careful consideration is given to the derivation of the Poisson
bracket between the alignment vector ${\bf{P}}$ and the momentum
density ${\bf{g}}$ and that between the director ${\bf{n}}$ and
${\bf{g}}$.  A similar approach was taken in recent work on
deriving the low-frequency hydrodynamics of nematic liquid
crystals and nematic polymers~\cite{Kamien2000, Lubensky2003}. In
section IV and V, we obtain the dynamical equation for the
director ${\bf{n}}$ and the alignment vector ${\bf{P}}$, as well
as for the conserved densities.  These equations contain new terms
violating the $\mathcal{Z}_2$-symmetry of the nematic phase.
Finally, in section VI we examine  the hydrodynamic modes and
demonstrate the splay instability induced by the new
$\mathcal{Z}_2$-odd terms.

\section{Definition of Microscopic Field Variables}

We start by defining the microscopic quantities of interest.  The
system is composed of $N$ identical liquid crystal molecules,
indexed by $\alpha$.  Each molecule consists of $s$ atoms treated
as point particles, with $m^{\mu}$ and $q^{\mu}$ the mass and the
charge of the $\mu$-th atom, respectively.  We denote by
${\bf{r}}^{\alpha\mu}$ the position of the $\mu$-th atom in the
$\alpha$-th molecule.  The position of the center of mass of the
$\alpha$-th molecule is given by
\begin{eqnarray}
{\bf{r^{\alpha}}}&=&\frac{\sum_{\mu}
m^{\mu}{\bf{r^{\alpha\mu}}}}{\sum_{\mu} m^{\mu}}\;,
\end{eqnarray}
and $\sum_{\mu}m^{\mu}=m^0$ is the molecular mass.  We assume that the molecules are neutral,
\begin{eqnarray}
\sum_{\mu}q^{\mu}&=&0\;,
\end{eqnarray}
but possess a non-vanishing dipole moment ${\bf{d^{\alpha}}}$, given by
\begin{eqnarray}
{\bf{d^{\alpha}}}&=&\sum_{\mu}q^{\mu}{\bf{r^{\alpha\mu}}}\equiv
d^\alpha  {\bf {\hat{\nu}}}^{\alpha}\;,
\end{eqnarray}
where ${\bf \hat{\nu}}^\alpha$ is a unit vector along the
molecular axis,
\begin{eqnarray}
{\bf {\hat{\nu}}}^\alpha&=&\frac{\sum_{\mu}q^{\mu}{\bf{r}}^{\alpha\mu}}
{\left\vert\sum_{\mu}q^{\mu}{\bf{r}}^{\alpha\mu}\right\vert}\;,
\end{eqnarray}
and
\begin{eqnarray}
d^{\alpha}={\bf {\hat{\nu}}}^{\alpha}\cdot{\bf{d}}^{\alpha}\;.
\label{d1}
\end{eqnarray}
The microscopic mass density, ${\hat{\rho}}({\bf{r}})$, momentum
density, ${\hat{{\bf{g}}}}({\bf{r}})$, and dipole moment density,
${\hat{{\bf{d}}}}({\bf{r}})$ are defined in the usual way as
\begin{eqnarray}
{\hat{\rho}}({\bf{r}})&=&\sum_{\alpha,\mu}m^{\mu}\delta({\bf{r}}-{\bf{r^{\alpha\mu}}})\;,\\
{\bf{\hat{g}(r)}}&=&\sum_{\alpha,\mu}{\bf{p^{\alpha\mu}}}\,\delta({\bf{r}}-{\bf{r^{\alpha\mu}}})\;,\\
{\bf{{\hat{d}}(r)}}&=&\sum_{\alpha}{\bf{d^{\alpha}}}\,\delta({\bf{r}}-{\bf{r^{\alpha}}})\;,
\label{d2}
\end{eqnarray}
where ${\bf p^{\alpha\mu}}$ is the momentum of the $\mu$-th atom
on the $\alpha$-th molecule. The macroscopic mean variables
describing the dynamics of equilibrium fluctuations in the system
are obtained from the microscopic ones after coarse-graining,
$\rho({\bf{r}})=\left[{\hat{\rho}}({\bf{r}})\right]_c$,
${\bf{g}}({\bf{r}})=\left[{\hat{\bf{g}}}({\bf{r}})\right]_c$,
${\bf{d}}({\bf{r}})=\left[{\hat{\bf{d}}}({\bf{r}})\right]_c$, as
described, for instance, in Refs.~\cite{LubenskyText} and
\cite{Lubensky2003}.

The macroscopic coarse-grained fields, denoted by $\Phi_{a}({\bf{r}},t)$, are hydrodynamic or
quasi-hydrodynamical variables whose characteristic decay times
are much longer than the underlying microscopic time scales and
diverge in the limit of long wavelength.  They are either
conserved variables or variables associated with broken symmetries
of the system.  They evolve in time according
to~\cite{LubenskyText,Lubensky2003}
\begin{eqnarray}
\frac{\partial \Phi_{a}({\bf{r}},t)}{\partial t}&=&V^a({\bf{r}})
-\Gamma_{ab}\frac{\partial {\mathcal{H}}}{\partial \Phi_{b}({\bf{r}})}\;,
\label{equation}
\end{eqnarray}
where $\Gamma_{ab}$ is the dissipative tensor, ${\mathcal{H}}$ is the systems Hamiltonian,  and $V^a$ is the reactive term or nondissipative velocity given by
\begin{eqnarray}
V_{a}({\bf{r}})&=&-\int d^3r' P_{ab}({\bf{r,r'}})\frac{\delta
{\mathcal{H}}}{\delta \Phi_{b}({\bf{r'}})}\;. \label{V}
\end{eqnarray}
The Einstein summation convention on repeated indices is understood and
\begin{eqnarray}
P_{ab}({\bf{r,r'}})&=&\{\Phi_{a}({\bf{r}}),\Phi_{b}({\bf{r}})\}\nonumber\\
&=&-P_{ba}({\bf{r',r}})
\end{eqnarray}
denotes the Poisson bracket of the coarse-grained variables. This
is defined as
\begin{eqnarray}
\{\Phi_{a}({\bf{r}}),\Phi_{b}({\bf{r'}})\}&=&\left[\{{\hat{\Phi}}_{a}({\bf{r}}),{\hat{\Phi}}_{b}({\bf{r'}})\}\right]_c\;,
\end{eqnarray}
with
\begin{eqnarray}
\{A({\bf{r}}),B({\bf{r'}})\}&=&\sum_{\alpha,\mu}\sum_i\left(\frac{\partial A({\bf{r}})}{\partial p^{\alpha\mu}_{ i}}\,\frac{\partial B({\bf{r'}})}{\partial r^{\alpha\mu}_{ i}}\right.\nonumber\\
&&-\left.\frac{\partial A({\bf{r}})}{\partial r^{\alpha\mu}_{
i}}\, \frac{\partial B({\bf{r'}})}{\partial p^{\alpha\mu}_{
i}}\right)\;. \label{PBdefinition}
\end{eqnarray}
The kinetic tensor $\Gamma_{\mu\nu}$ must be symmetric and
$\partial\Phi_{\mu}({\bf{r},t})/\partial t$ and $\delta{\mathcal{
H}}/\delta\Phi_{\nu}$ must have opposite signs under time
reversal. The coarse-grained dynamical equations (\ref{equation})
can be rigorously derived from microscopic
principles~\cite{Zwanzig, Kawasaki, MFS}.

Polar order is described by the alignment vector ${\bf{P(r)}}$,
defined as
\begin{eqnarray}
{\bf{P}}({\bf{r}})&\equiv&\sum_{\alpha}\,{\bf
{\hat{\nu}}}^{\alpha}\delta({\bf{r-r_{\alpha}}})\;. \label{d4}
\end{eqnarray}
The alignment vector ${\bf{P}}({\bf{r}})$ embodies the
fluctuations in the dipoles' orientation.  We also write
\begin{eqnarray}
{\bf{P}}({\bf{r}})&\equiv&\rho({\bf
r})S({\bf{r}}){\bf{n}}({\bf{r}})\;. \label{d5}
\end{eqnarray}
The scalar field $S({\bf{r}})$ encodes information on the degree
of local alignment of the dipoles, i.e., $S({\bf{r}})$ is the
order parameter for the isotropic-ferroelectric nematic phase
transition.  The unit vector ${\bf{n}}({\bf{r}})$ is the
macroscopic director.   As previously mentioned, the vectorial
nature of ${\bf{P(r)}}$ is in sharp contrast to the tensorial
character of the usual nematic order parameter, $Q_{ij}(r)$. This
mathematical distinction leads to new terms in the free energy of
a polar fluid that are not allowed for a nematic liquid crystal.
\\
\\
\section{Free energy and stability of the polarized phase}
Before proceeding with the derivation of the hydrodynamic
equations for the polarized state, it is instructive to discuss
the coarse-grained free energy of a polar fluid. This is given by
\begin{eqnarray}
F&=&F_p+F_n\;,\\
\label{Fpn}
F_{p}&=&\int d^3r \,\,\,\,B\frac{\delta\rho({\bf{r}})}{\rho_0}\nabla\cdot{\bf{n(r)}}+\ldots\;,\nonumber\\
\\
F_{n}&=&\frac{1}{2}\int d^3r \,\,\,\,\left\{C_1\Big(\frac{\delta\rho}{\rho_0}\Big)^2+C_2(\nabla\rho)^2+\ldots\right. \nonumber\\
&&+K_1(\nabla\cdot{\bf{n}})^2+K_2({\bf{n\cdot\nabla\times n}})^2\nonumber\\
&&\left.+K_3({\bf{n\times\nabla\times n}})^2+\ldots\right\}\;,
\label{Fn}
\end{eqnarray}
where $\delta\rho({\bf r})=\rho({\bf r})-\rho_0$ is the deviation
of the local density from its equilibrium value, $\rho_0$. The
contribution $F_n$ is the free energy of a nematic liquid crystal
which contains only terms symmetric with respect to
${\bf{n}}\rightarrow {\bf{-n}}$, with $K_1, K_2$ and $K_3$  the
usual Frank constants for splay, twist and bend deformations,
respectively. The contribution $F_{p}$ contains additional terms
that break that symmetry and are only allowed in a fluid that is
macroscopically polar.  In general a term
$\sim\nabla\cdot{\bf{n(r)}}$ is also allowed in the free energy.
This will give rise to a surface term that will favor a splay
distortion of the molecules (corresponding to a non-zero
$\nabla\cdot{\bf{n(r)}}$). In the absence of surface
stabilization, this would always destroy the polar order. In this
work we assume that there is sufficient surface stabilization to
suppress the surface term and to prevent the associated splay.

In addition to a possible surface instability, it is apparent from
the form of the free energy that a polarized liquid crystal can
also exhibit a {\em bulk} instability to splay deformations. The
latter is associated with the term
$\sim\delta\rho({\bf{r}})\nabla\cdot{\bf{n(r)}}$ that cannot be
eliminated by surface stabilization. The existence of this
instability can be understood by rewriting the terms in the free
energy involving density fluctuations and splay deformations of
the director  as
\begin{widetext}
\begin{eqnarray}
\int \frac{d^3r}{2}
\Big\{C_1\Big(\frac{\delta\rho}{\rho_0}\Big)^2+K_1(\nabla_\perp\cdot\delta{\bf
n})^2+2B\frac{\delta\rho}{\rho_0} \bm\nabla_\perp\cdot\delta{\bf
n}\Big\}= \int \frac{d^3r}{2}
\Big\{\frac{C_1}{\rho_0^2}\big[\delta\rho+\frac{B\rho_0}{C_1}\bm\nabla_\perp\cdot\delta{\bf
n}\big]^2
+\big[K_1-\frac{B^2}{C_1}\big](\bm\nabla_\perp\cdot\delta{\bf
n})^2\Big\}\;,
\end{eqnarray}
\end{widetext}
where $\bm\nabla=(\bm\nabla_\perp,\partial_z)$. It is then clear
that the effect of the coupling of density fluctuations to splay
yield a downward renormalization of the splay elastic constant,
$K_1$. For $|B|>\sqrt{K_1C_1}$ splay fluctuations becomes
energetically favorable and will clearly destabilize the ordered
state. The instability is suppressed if the fluid is
incompressible since the renormalization of the Frank constant
$K_1$ vanishes when $C_1\rightarrow\infty$. The stable ground
state for $|B|>\sqrt{K_1C_1}$ is expected to be characterized by
 director configurations that are spatially inhomogeneous in the
$x$ direction, perpendicular
 to the broken symmetry direction, with associated spatial
 structures
in the density. The precise director configurations will depend on
boundary effects, as well as on applied external fields (electric
field in the case of ferroelectric order). The dynamics of the
polarized state discussed below will of course reflect the
existence of this instability, which will lead to the growth of
the hydrodynamic mode associated with splay fluctuations for
$|B|>\sqrt{K_1C_1}$.

In order to examine the possible relevance of the splay
instability to experimental systems we need to estimate the
various elastic constants. We consider a fluid of rods of linear
size $l$ interacting via a short-range repulsive interaction of
strength $U$. The rods carry a permanent electric dipole moment of
magnitude $d$, which yields a dipolar coupling of strength
$U_D\sim k_c d^2/r^3$ between two molecules at distance $r$, where
$k_c$ is the Coloumb's law constant. The bulk compressional
modulus $C_1$ has dimensions of energy density and its size is
controlled by the repulsive part of the interaction, with $C_1\sim
U/l^3$. The splay Frank coefficient has dimensions of
energy/length and can be estimated as $K_1\sim U/l$. This yields
an estimate for the critical value $B_0=\sqrt{C_1K_1}$ where the
splay instability occurs: $B_0\sim U/l^2$. The elastic constant
$B$ that couples density and splay fluctuations has dimensions of
${\rm energy}/({\rm length})^2$ and is controlled by interactions that favor alignment of the liquid crystal molecules (dipole-dipole interaction in the case of ferroelectric order).  Clearly the elastic constant $B$ would vanish if the molecules are just rods as in the nematic case.  We estimate $B\sim U_D/l^2$, where
$U_D$ is the strength of the dipolar interaction (for dipoles at a
distance $r\sim\rho_0^{-1/3}$, we estimate $U_D\sim k_c
d^2\rho_0$). The splay instability will then occur for $U_D\sim
U$.  Taking $\rho_0 \sim 10^{-27} m^{-3}$ and $d \sim$ 2 Debye
\cite{Dipole Moment Estimate}
 we estimate $U_d \sim 2.5$ meV. This is smaller than $U \sim 0.1$eV, so we see that a
stable ferroelectric phase of a polar liquid crystal is, in
principle, possible. Of course, this assumes appropriate surface
stabilization as discussed above.

\section{Poisson brackets for a polar liquid crystal}
We now return to the derivation of hydrodynamics via the
Poisson-bracket method. Using Eq.(\ref{PBdefinition}), it is
straightforward to evaluate the Poisson-bracket  relations between
the various hydrodynamic fields.  Our goal is to obtain equations
describing the dynamics at long wavelengths.  We will therefore
only keep terms of lower order in the gradients of the
hydrodynamic fields and expand the $\delta$-function as
\begin{eqnarray}
\delta({\bf{r}}-{\bf{r^{\alpha\mu}}})&=&\delta({\bf{r}}-{\bf{r^{\alpha}}}-\Delta{\bf{r^{\alpha\mu}}})\nonumber\\
&=&\delta({\bf{r}}-{\bf{r^{\alpha}}})-\Delta{{r^{\alpha\mu}_{ k}}}\nabla_k\delta({\bf{r}}-{\bf{r^{\alpha}}})\nonumber\\
&&+O(\nabla^2)\;.\nonumber\\
\label{delta}
\end{eqnarray}
where $\Delta{\bf{r^{\alpha\mu}}}={\bf{r}}^{\alpha\mu}-{\bf{r^{\alpha}}}$.
The required non-vanishing Poisson-brackets are:
\begin{eqnarray}
\label{pg}
\{\rho({\bf{r}}),g_i({\bf{r'}})\}&=&\nabla_i\delta({\bf{r-r'}})\rho({\bf{r'}})\;,\\
\label{gp}
\{g_i({\bf{r}}),g_j({\bf{r'}})\}&=& g_i({\bf
r'})\nabla_j\delta({\bf{r-r'}})\nonumber\\
&&-\nabla'_i\left[\delta({\bf{r-r'}}) g_j({\bf r'})\right]\;,\\
\{P_i({\bf{r}}),g_j({\bf{r'}})\}&=&\nabla_{j}\left[P_i({\bf{r}})\delta({\bf{r-r'}})\right]\nonumber\\
&&-\delta_{ij}P_k({\bf{r}})\nabla_k\delta({\bf{r}}-{\bf{r'}})\nonumber\\
&&+\lambda_{ijk}({\bf r})\nabla_k\delta({\bf{r}}-{\bf{r}})\;,
\label{dg0}
\end{eqnarray}
where
\begin{eqnarray}
{\hat{\lambda}}_{ijk}&=&\sum_{\alpha}\bf{{\hat{\nu}}}^{\alpha}_i{\bf {\hat{\nu}}}^{\alpha}_j{\bf {\hat{\nu}}}^{\alpha}_k
\delta({\bf{r-r_{\alpha}}})
\end{eqnarray}
is a symmetric third-order tensor that depends on the degree of
molecular alignment.  We assume that its coarse-grained
counterpart $\lambda_{ijk}$ depends only on the alignment vector ${\bf{P(r)}}$.
Quite generally, $\lambda_{ijk}$ can then be written as
\begin{eqnarray}
\lambda_{ijk}&=&\lambda_1\,\left(\delta_{ij}P_k+\delta_{jk}P_i+\delta_{ik}P_j\right)
+\lambda_2~\frac{P_iP_jP_k}{P^2}\;,\nonumber\\
\label{PPP}
\end{eqnarray}
where $P({\bf r})={\bf n}\cdot{\bf P}=\rho S$. The two
coefficients $\lambda_1$ and $\lambda_2$ are not independent. By
contracting both sides of Eq.~(\ref{PPP}) with suitble
combinations of $\delta_{ij}$ and $n_k$ we obtain
\begin{eqnarray}
n_k\lambda_{iik}=\rho S=(4\lambda_1+\lambda_2)\rho S\;,
\end{eqnarray}
\begin{eqnarray}
n_in_jn_k\lambda_{ijk}&=&\sum_{\alpha}\left({\bf
{\hat{\nu}}}_{\alpha }\cdot{\bf{n}}\right)^3\,
\delta({\bf{r}}-{\bf{r}}_{\alpha})\nonumber\\
&=&(3\lambda_1+\lambda_2)\rho S\;.
\end{eqnarray}
The two coefficients can then be expressed in terms of  single
microscopic quantity, $\lambda$, given by
\begin{eqnarray}
\lambda&=&\frac{1}{\rho({\bf r})S({\bf
r})}\Big[\sum_{\alpha}\left({\bf {\hat{\nu}}}_{\alpha
}\cdot{\bf{n}}\right)^3\,
\delta({\bf{r}}-{\bf{r}}_{\alpha})\Big]_c\;,
\end{eqnarray}
as
\begin{eqnarray}
&&\lambda_1=\frac{1-\lambda}{2}\;,\\
&&\lambda_2= \frac{5\lambda-3}{2}\;.
\end{eqnarray}
Equation.~(\ref{PPP}) can then be rewritten in terms of $\lambda$
as
\begin{eqnarray}
\lambda_{ijk}&=&\frac{1-\lambda}{2}\Big[\delta^T_{ij}P_k+\delta^T_{ik}P_j+\delta^T_{jk}P_i\Big]
+\lambda~ n_j n_k P_i\;.
\end{eqnarray}
where $\delta^T_{ij}$ is an operator that projects transverse to
the director,
\begin{eqnarray}
\delta^T_{ij}({\bf r})&=&\delta_{ij}-n_i({\bf{r}})n_j({\bf{r}})\;.
\end{eqnarray}

The director field itself has no microscopic definition in terms
of the canonical coordinates and momenta of the individual atoms.
Its Poisson-brackets must therefore be obtained indirectly from
those of the alignment vector ${\bf{P}}$. Using the definition of
the director given in Eq.~(\ref{d5}) and the chain rule of
derivatives we can write
\begin{eqnarray}
\left\{P_i({\bf r}),g_j({\bf r'})\right\}&=&\rho({\bf r})S({\bf
r})\left\{n_i({\bf r}),g_j({\bf r'})\right\}\nonumber\\
&&+n_i({\bf r})\left\{\rho({\bf r})S({\bf r}),g_j({\bf
r'})\right\}\;.
\end{eqnarray}
Since
$n_i({\bf{r}})\left\{n_i({\bf{r}}),\Phi_{\mu}({\bf{r'}})\right\}=0$,
for any field $\Phi_{\mu}$, we obtain
\begin{eqnarray}
\{n_i({\bf{r}}),g_j({\bf{r'}})\}&=&\frac{1}{\rho({\bf
r})S({\bf{r}})} \delta^T_{ik}({\bf
r})\left\{P_k({\bf{r}}),g_j({\bf{r'}})\right\}\;. \label{ng3}
\end{eqnarray}

\section{hydrodynamic equations for the Polarized Phase}

In this section we derive the hydrodynamic equations in the
polarized phase.  The order parameter $S({\bf{r}})$ is assumed to
be finite and constant in the following discussion. To calculate
the reactive part of the hydrodynamic equations defined in
Eq.~(\ref{V}) in terms of the Poisson brackets, we need the
coarse-grained Hamiltonian for the system
\begin{eqnarray}
\label{H} {\mathcal{H}}[\rho,{\bf{g,n}}]&=&\int d^3 r\,\,\,\,\frac{{\bf{g}}^2({\bf{r}})}{2\rho({\bf{r}})}+F[\rho,{\bf{n}}]\;.
\end{eqnarray}
It consists of a kinetic part and a free energy $F[\rho,{\bf{n}}]$
that depends on the specific system in question and will be
specified below.

The nondissipative velocity for the mass-density equation is immediately found to be
\begin{eqnarray}
V^{\rho}&=&-\nabla_ig_i({\bf{r}})\;.
\end{eqnarray}

The determination of the non-dissipative term for the
momentum-density equation is straightforward albeit more
tedious.  Using Eqs. (\ref{pg}) and (\ref{ng3}), we obtain
\begin{eqnarray}
V^{{\bf{g}}}_i&=&-\nabla_j\frac{g_i({\bf{r}})g_j({\bf{r}})}{\rho}-\rho({\bf{r}})\nabla_i\frac{\delta F}{\delta\rho({\bf{r}})}\nonumber\\
&&-n_k\nabla_i\left[\delta^T_{jk}\frac{\delta F}{\delta n_j({\bf{r}})}\right]\nonumber\\
&&+\nabla_k\left[n_k\delta^T_{ij}\frac{\delta F}{\delta n_i({\bf{r}})}\right]\nonumber\\
&&-\nabla_k\left[\frac{\delta^T_{jl}\,\lambda_{ikl}}{\rho({\bf{r}})S({\bf{r}})}
\frac{\delta F}{\delta n_j({\bf{r}})}\right]\;.
\end{eqnarray}
Similarly for the director, we obtain using Eq.(\ref{ng3}),
\begin{eqnarray}
V^{\bf{n}}_{i}&=&-v_j\nabla_jn_i+\omega_{ij}n_j+\lambda\delta^T_{ik}u_{kj}n_j\;,
\label{vn}
\end{eqnarray}
where $v_i=\delta {\mathcal{H}}/\delta g_i$ is the velocity and $u_{ij}=\frac 12(\nabla_iv_j+\nabla_jv_i)$ is
the strain-rate tensor, while $\omega_{ij}=\frac
12(\nabla_iv_j-\nabla_jv_i)$ is the vorticity. Only the part of
the tensor $\lambda_{ijk}$ that is transverse to the director
enters in the hydrodynamic equations. This is given by
\begin{eqnarray}
\delta^T_{jl}({\bf{r}})\lambda_{ikl} ({\bf{r}})
=&&\frac{1-\lambda}{2}\rho({\bf r})S({\bf{r}})\nonumber\\
&&\times\left[\delta^T_{ij}({\bf{r}})n_k({\bf{r}})
+\delta^T_{jk}({\bf{r}})n_i({\bf{r}})\right]\;.\nonumber\\
\end{eqnarray}

We now turn our attention to the dissipative part of Eq.
(\ref{equation}).  Since the mass-conservation is exact, there is
no dissipative term allowed in the dynamical equation for the
mass-density.   In the case of unbroken Galilean invariance, the
time derivative of ${\bf{g}}$ can only couple to the gradients of
velocity.  As in the nematic
case~\cite{Lubensky2003}, we introduce the tensor of viscosities
$\eta_{ijkl}$ and the viscous stress tensor $\sigma^V_{ij}$, with
the following properties:
\begin{eqnarray}
\nabla_j\eta_{ijkl}\nabla_kv_l&=&\nabla_j\sigma^V_{ij}\;,
\end{eqnarray}
where
\begin{eqnarray}
\sigma^V_{ij}&=&\alpha_1n_in_jn_kn_lu_{kl}+\alpha_4u_{ij}\nonumber\\
&&+\frac{\alpha_5+\alpha_6}{2}(n_iu_{jk}+n_ju_{ik})n_k\nonumber\\
&&+\rho_1\delta_{ij}u_{kk}+\rho_2(\delta_{ij}n_kn_lu_{kl}+u_{kk}n_in_j)\;.
\label{SigmaV}
\end{eqnarray}
To the lowest order in gradients, there is no difference in the
expression of $\sigma^V_{ij}$ between the polar and nematic cases.
Likewise, as in a nematic liquid crystal, the time derivative
$\partial {\bf{n}}/\partial t$ can only couple to a term of the
form
\begin{eqnarray}
-\gamma\delta^T_{ij}\frac{\delta F}{\delta n_j}\;,
\end{eqnarray}
where $\gamma^{-1}$ is a rotational friction.

Thus, collecting the reactive and dissipative terms, we arrive at
the following hydrodynamic equations for a liquid crystal with
polar orientational order
\begin{eqnarray}
\frac{\partial\rho}{\partial t}&=& -{\bf{\nabla\cdot
g({\bf{r}})}}\;, \label{pnn}
\end{eqnarray}
\begin{eqnarray}
\frac{\partial g_i}{\partial t}&=&-\nabla_j\frac{g_i({\bf{r}})g_j({\bf{r}})}{\rho}
-\rho({\bf{r}})\nabla_i\frac{\delta F}{\delta\rho({\bf{r}})}\nonumber\\
&&+(\nabla_in_j)\left(\frac{\delta F}{\delta n_j}\right)
+\frac{1+\lambda}{2}\nabla_k\left[n_k\delta^T_{ij}\frac{\delta F}{\delta n_j}\right]\nonumber\\
&&-\frac{1-\lambda}{2}\nabla_k\left[n_i\delta^T_{jk}\frac{\delta
F}{\delta n_j}\right]+\nabla_j\sigma^V_{ij}\;, \label{gnn}
\end{eqnarray}
\begin{eqnarray}
\frac{\partial n_i}{\partial t}&=&-v_j\nabla_jn_i-\omega_{ij}n_j+\lambda~\delta^T_{ik}u_{kj}n_j\nonumber\\
&&-\gamma\delta^T_{ij}\frac{\delta F}{\delta n_j}\;. \label{nnn}
\end{eqnarray}
We note that the hydrodynamic equations for the polarized phase
have the same form as those for a conventional nematic, as given
for instance in~\cite{Lubensky2003}, although of course with a
different microscopic expression for $\lambda$.  The only
differences between the two sets of equations come from
differences in the free energy.

Using the form for the free energy given in Eqs.~(\ref{Fn}) and
(\ref{Fpn}), the hydrodynamic equations of a polar liquid crystal
are given by the continuity equation (\ref{pnn}) and
\begin{eqnarray}
\partial_tn_i({\bf{r}})
&=&-v_j({\bf{r}})\nabla_j n_i({\bf{r}})+\omega_{ij} n_j+\lambda\delta^T_{ik}u_{kj}n_j\nonumber\\
&&+\gamma\delta^T_{ij}\frac{B}{\rho_0}\nabla_j\rho({\bf{r}})-\gamma\delta^T_{ij}\frac{\delta
F_n}{\delta n_j}\;, \label{nnt}
\end{eqnarray}
\begin{eqnarray}
\partial_t g_i({\bf{r}})&=&-\nabla_j\frac{g_i({\bf{r}})g_j({\bf{r}})}{\rho}
-B\frac{\rho({\bf{r}})}{\rho_0}\nabla_i\nabla_jn_j({\bf{r}})\nonumber\\
&&-\frac{B}{\rho_0}(\nabla_in_j)(\nabla_j\rho)-\frac{1+\lambda}{2\rho_0}B\nabla_k\left[\delta^T_{ij}n_k\nabla_j\rho\right]\nonumber\\
&&+\frac{1-\lambda}{2\rho_0}B\nabla_k\left[\delta^T_{jk}n_i\nabla_j\rho\right]\nonumber\\
&&-\rho({\bf{r}})\nabla_i\frac{\delta F_n}{\delta\rho({\bf{r}})}
+(\nabla_in_j)\left(\frac{\delta F_n}{\delta n_j}\right)\nonumber\\
&&+\frac{1+\lambda}{2}\nabla_k\left[n_k\delta^T_{ij}\frac{\delta F_n}{\delta n_j}\right]\nonumber\\
&&-\frac{1-\lambda}{2}\nabla_k\left[n_i\delta^T_{jk}\frac{\delta
F_n}{\delta n_j}\right]+\nabla_j\sigma^{V}_{ij}\;,
\label{ggg}
\end{eqnarray}
where
\begin{eqnarray}
\label{dFp}
\frac{\delta F_{n}}{\delta \rho}&=&C_1\left(\frac{\delta\rho}{\rho_0^2}\right)-C_2\nabla^2\rho\;,\\
\frac{\delta F_n}{\delta n_j}&=&-K_1\nabla_j(\nabla\cdot{\bf{n}})+K_2\left(\nabla\times(\nabla\times{\bf{n}})\right)_j\nonumber\\
&&-(K_2-K_3)n_s(\nabla_sn_k)(\nabla_jn_k)\nonumber\\
&&+(K_2-K_3)\nabla_b\left(n_bn_s(\nabla_sn_j)\right) \;.
\label{dFn}
\end{eqnarray}
The terms in the hydrodynamic equations that are unique to a fluid
with macroscopic polar order are those proportional to the elastic
constant $B$. These are forbidden by symmetry in the hydrodynamic
equations of a nematic.

\section{Dynamics of the alignment vector}

In the more general case where $S({\bf{r}})$ is not a constant in
Eq. (\ref{d5}), we need to consider the full order parameter of
the alignment vector ${\bf{P}}$.  The hydrodynamic equations for
the alignment vector ${\bf{P}}$ are
\begin{eqnarray}
\label{ptt}
\frac{\partial}{\partial t} \rho({\bf{r}})&=&-{\bf{\nabla\cdot g}}\;,\\
\label{ttt}
\frac{\partial}{\partial t} P_i({\bf{r}})&=&-v_j({\bf{r}})\nabla_jP_i({\bf{r}})+P_j({\bf{r}})\nabla_jv_i({\bf{r}})\nonumber\\
&&-P_i({\bf{r}})(\nabla\cdot{\bf{v}})-\lambda_{ijk}\nabla_kv_j({\bf{r}})-{\gamma'}\frac{\delta F}{\delta P_i}\;,\\\label{Ptt}
\frac{\partial}{\partial t} g_i({\bf{r}})&=&-\nabla_j\frac{g_i({\bf{r}})g_j({\bf{r}})}{\rho}-\rho({\bf{r}})\nabla_i\frac{\delta F}{\delta\rho({\bf{r}})}\nonumber\\
&&+\nabla_k\left[P_k({\bf{r}})\frac{\delta F}{\delta P_i({\bf{r}})}\right]-P_j({\bf{r}}\left[\nabla_i\frac{\delta F}{\delta P_j({\bf{r}})}\right]\nonumber\\
&&-\nabla_k\left[\lambda_{ijk}\,\frac{\delta F}{\delta
P_j({\bf{r}})}\right]-\nabla_j\sigma^V_{ij}\;. \label{gtt}
\end{eqnarray}
%
The equation for the director can be obtained from Eq.~
(\ref{ttt}) by using the relation
\begin{eqnarray}
\partial_tn_i({\bf{r}})&=&\frac{1}{\rho({\bf r})S({\bf{r}})}\delta^T_{ik}({\bf r})\partial_tP_k({\bf{r}})\;.
\end{eqnarray}

\section{hydrodynamic modes in the polarized phase}

In the isotropic phase the only hydrodynamic fields are the four
conserved densities: number density and three components of the
momentum (energy fluctuations are not considered here). The
corresponding four hydrodynamic modes are those of a conventional
isotropic liquid: two propagating sound wave describing the decay
of density and longitudinal momentum fluctuations, and two
diffusive shear modes, describing the decay of the two tranverse
components of the momentum density. We now consider the long
wavelength, low frequency dynamics in a uniformly polarized phase,
characterized by uniform equilibrium values $\rho_0$ of the
density, ${\bf v}_0=0$ of the flow velocity and ${\bf n}_0={\bf
{\hat{z}}}$ of the director field. In order to evaluate the
hydrodynamic modes of the polarized state, we expand the
hydrodynamic equations to linear order in the fluctuations of
density, $\delta\rho = \rho-\rho_0$, momentum
$\delta{\bf{g}}\simeq\rho_0{\bf{v}}$, and director ${\bf{\delta
n=n-{\hat{z}}}}$ about their equilibrium values (to lowest order
$\bf {\hat{z}}\cdot {\bf {\delta n}}=0$).  We will ignore any variation in temperature in our consideration.
We consider wave-like fluctuations whose space and time dependence
is of the form $\exp[i\,{\bf{q\cdot r}}-i\omega t]$ and choose the
coordinate system so that the wavevector ${\bf q}$ lies in the
$(xz)$ plane, i.e., ${\bf{q}}=(q_x,q_z)$ and $q_y=0$. We then
project velocity and director fluctuations along the axes of an
orthogonal coordinate system defined by the unit vectors
$\big({\bf\hat{z}},{\bf{\hat{q}}}_x,{\bf\hat{z}}\times
{\bf{\hat{q}}}_x\big)$,  with ${\bf\hat{q}_x}= q_x/q$.  The
equations for velocity and director fluctuations transverse to
both the direction of polarization ${\bf\hat{z}}$ and
${\bf\hat{q}_x}$ defined as
\begin{eqnarray}
&&v_{\bf q}^{y}=({\bf\hat{z}}\times{\bf\hat{q}_x})\cdot{\bf v_{q}}\;,\\
&&\delta n_{\bf
q}^{y}=({\bf\hat{z}}\times{\bf\hat{q}_x})\cdot\delta{\bf n_{q}}\;,
\end{eqnarray}
decouple from the others
and are given by
\begin{eqnarray}
&&\partial_t\delta n_{\bf q}^{y}=-\gamma K_T({\bf\hat{q}})q^2\delta
n_{q_y}-i\frac{1-\lambda}{2}q_zv_{\bf q}^{y}\;,\\
&&\partial_t
v_{\bf q}^{y}=-\nu_T({\bf\hat{q}})q^2v_{q_y}+i\frac{1+\lambda}{2\rho_0}K_T({\bf\hat{q}})q^2q_z\delta
n_{\bf q}^{y}\;,
\end{eqnarray}
where ${\bf\hat{q}}=(q_x/q,q_z/q)$ and
\begin{eqnarray}
K_T({\bf \hat{q}})&=&\big[K_2\,q_x^2+K_3q_z^2\big]/q^2\;,\\
\nu_T({\bf
\hat{q}})&=&\frac{\alpha_4}{2\rho_0}+\frac{\alpha_5+\alpha_6}{4\rho_0}~\frac{q_z^2}{q^2}\;.
\end{eqnarray}
The transverse eigenvalues are identical to those of nematic
liquid crystals and are given by
\begin{eqnarray}
i\,\omega^T_\pm&=&\frac{q^2}{2}\big[\nu_T({\bf\hat{q}})+\gamma
K_T({\bf\hat{q}})\big]\nonumber\\
&&\pm\frac{q^2}{2}\sqrt{\big[\nu_T({\bf\hat{q}})-\gamma
K_T{\bf\hat{q}})\big]^2+\hat{q}_z^2\frac{1-\lambda^2}{\rho_0}K_T({\bf\hat{q}})}\;.\nonumber\\
\end{eqnarray}
These modes are always diffusive and describe the coupled decay of
vorticity and twist fluctuations. For $q_z=0$ the transverse modes
decouple and are simply given by
\begin{eqnarray}
&&i\,\omega^T_{n}=\gamma K_2q_x^2\;,\\
&&i\,\omega^T_{v}=\frac{\alpha_4}{2\rho_0}q_x^2\;,
\end{eqnarray}
where $\omega^T_{n}$ describes the diffusion of fluctuations in
the local twist and $\omega^T_{v}$ controls the decay of
fluctuations in the vorticity.

The remaining four hydrodynamic fluctuations, describing the
dynamics of density $\delta\rho_{\bf{q}}$, velocity and director
components longitudinal to ${\bf\hat{q}_x}$, $v_{\bf
q}^{x}={\bf\hat{q}_x}\cdot{\bf v_q}$ and $\delta n_{\bf
q}^{x}={\bf\hat{q}_x}\cdot\delta{\bf n_q}$, and velocity component
along the direction of polarization, $v_{\bf q}^{z}$ are coupled.
The linearized equations are given by
\begin{widetext}
\begin{eqnarray}
\label{rhoeqn}\partial_t\delta\rho_{\bf{q}}&=&-i\rho_0q_x v_{\bf q}^{x}-i\rho_0q_z v_{\bf q}^{z}\;,\\
\label{nxeqn}\partial_t\, \delta {{n}}_{\bf q}^{x}&=&-\gamma
K_L({\bf \hat{q}})q^2\delta{{n}}_{\bf q}^{x}+i\gamma
\frac{B}{\rho_0}q_x\delta\rho_{\bf{q}}
-i\frac{1-\lambda}{2}q_zv_{\bf q}^{x}+i\frac{1+\lambda}{2}q_x
v_{\bf q}^{z}\;,\\
\label{vxeqn}\partial_t v_{\bf
q}^{x}&=&-\nu_L({\bf\hat{q}})q^2v_{\bf q}^{x}-\nu'_Lq_x q_zv_{\bf
q}^{z}-iq_x
\frac{C_1}{\rho_0^2}\delta\rho_{\bf{q}}+\frac{1+\lambda}{2\rho_0^2}Bq_zq_x\delta\rho_{\bf{q}}
+i\frac{1+\lambda}{2\rho_0}K_L({\bf\hat{q}})q_zq^2\delta n_{\bf q}^{x}\;,\\
\label{vzeqn}\partial_t v_{\bf
q}^{z}&=&-\nu_z({\bf\hat{q}})q^2v_{\bf q}^{z}-\nu'_Lq_x q_zv_{\bf
q}^{x}-iq_z
\frac{C_1}{\rho_0^2}\delta\rho_{\bf{q}}-\frac{1-\lambda}{2\rho_0^2}Bq_x^2\delta\rho_{\bf{q}}
-i\frac{1-\lambda}{2\rho_0}K_L({\bf\hat{q}})q_x q^2\delta n_{\bf
q}^{x}\;,
\end{eqnarray}
\end{widetext}
where
\begin{eqnarray}
&&K_L({\bf \hat{q}})=\big[K_1\,q_x^2+K_3q_z^2\big]/q^2\;,\\
&&\nu_{L}({\bf
\hat{q}})=\frac{\alpha_4+\rho_1}{\rho_0}~\frac{q_x^2}{q^2}
+\frac{2\alpha_4+\alpha_5+\alpha_6}{4\rho_0}~\frac{q_z^2}{q^2}\;,\\
&&\nu_{z}({\bf
\hat{q}})=\frac{\alpha_1+\alpha_4+\alpha_5+\alpha_6+\rho_1+2\rho_2}{\rho_0}~\frac{q_z^2}{q^2}\nonumber\\
&&\hspace{0.5in}+\frac{2\alpha_4+\alpha_5+\alpha_6}{4\rho_0}~\frac{q_x^2}{q^2}\;,\\
&&\nu'_L=\frac{\rho_1+\rho_2}{\rho_0}+\frac{2\alpha_4+\alpha_5+\alpha_6}{4\rho_0}\;.
\end{eqnarray}
The corresponding hydrodynamic equations for a nematic liquid
crystal are obtained from Eqs.~(\ref{rhoeqn}-\ref{vzeqn}) by
simply setting $B=0$. For comparison we recall that there are four
longitudinal hydrodynamic modes in a compressible nematic liquid
crystal: two propagating sound waves and two diffusive modes
describing the coupled decay of splay and velocity
fluctuations~\cite{SS74}.  The same four modes
are present in a polarized liquid crystal, but their decay rates
are modified in important ways by the new symmetry term breaking
term proportional to $B$.

We calculate the hydrodynamic frequencies up to $o(q^2)$ by
solving the eigenvalue problem perturbatively. As expected, we
find two propagating  and two purely diffusive modes. The two
propagating eigenmodes are the sound waves of the fluid and are
given by
\begin{eqnarray}
i\,\omega^{s}_\pm&\approx&\pm i\sqrt{\frac{C_1}{\rho_0}} q \nonumber\\
&&+\frac{q^2}{2} \left[ \nu_L \hat{q}_{x}^2 + \nu_z\hat{q}_z^2 + 2
\nu_L^\prime
 \hat{q}_{x}^2 \hat{q}_{z}^2 +
 \frac{\gamma B^2} {C_1} \hat{q}_{x}^2  \right].\nonumber\\
\end{eqnarray}
To lowest order in the wavevector the propagation speed is
proportional to $\sqrt{\frac{C_1}{\rho_0}}$. As in isotropic
liquids, it is determined entirely by the compressional modulus
$C_1$ of the fluid and receive corrections proportional to $B$.
The coupling to polar order described by the elastic constant $B$
increases the relaxation rate, but has no qualitative effect on
the structure of the sound waves. In the incompressible limit,
$C_1$ tends to infinity and all contributions to the dynamics of
the system from terms proportional to $B$ vanish.  This is evident
in Eq. (\ref{dFp}) as the $\mathcal{Z}_2$-odd term becomes a
surface term.

The eigenfrequencies of the two diffusive modes are given by
\begin{widetext}
\begin{eqnarray}
i\omega^{diff}_{\pm}&=&q^2\left[\frac{\gamma K({\bf
\hat{q}})+\mu({\bf \hat{q}})}{2}
\pm\frac12\sqrt{\left[\gamma K({\bf \hat{q}})-\mu({\bf \hat{q}})\right]^2+\frac{K({\bf \hat{q}})}{\rho_0}\left[1-\lambda^2\left({\hat{q}}_x^2-{\hat{q}}_z^2\right)^2\right]}\right]\nonumber\\
\end{eqnarray}
\end{widetext}
where
\begin{eqnarray}
K({\bf \hat{q}})&=& K_L({\bf \hat{q}})-\frac{B^2}{C_1}{\hat{q}}_x^2\;,\\
\mu({\bf \hat{q}})&=&\nu_{z}({\bf{\hat{q}}}){\hat{q}}_x^2
+\nu_{L}({\bf{\hat{q}}}){\hat{q}}_z^2-2\nu'_L{\hat{q}}_x^2{\hat{q}}_z^2\nonumber\\
&=&\Big[2\alpha_4+\alpha_5+\alpha_6+4\alpha_1\hat{q}_x^2\hat{q}_z^2\Big]/4\rho_0\;.
\end{eqnarray}
When $K({\bf \hat{q}})= 0$, the diffusive modes become unstable.
This is precisely the splay instability discussed on Sec. III by
examining the free energy of the system.

In conclusion, we have derived the dynamical equations for the
director {\bf{n}} and the alignment vector ${\bf{P}}$, starting
from the microscopic expressions of the conserved quantities and
carrying out rigorous coarse-graining via the Poisson-bracket
approach.  Our results are consistent with the equations for
nematic liquid-crystals derived earlier using similar
approaches~\cite{Kamien2000, Lubensky2003}.  Furthermore, we have
investigated the hydrodynamic modes of an ordered phase with
macroscopic polarization. We have found that the polarity of the
director field introduces a frustration in the system. As a
result,  a state of uniform local density and director orientation
can become unstable to the growth of long wavelength splay
fluctuations in a region of parameters. The splay instability
arises solely from the addition of $\mathcal{Z}_2$-odd terms in
the free energy, which are generically allowed on the basis of
symmetry considerations in any polar liquid. We therefore believe
that the existence of the instability is a general feature of any
polar fluid, regardless of the origin of the polarity.

It is our hope that the results reported in this paper will be
useful in providing a rigorous justification of the various
equations used previously in studying the ferroelectric
phase of polar liquid crystals.
It is our further hope that our results, presented here in
the context of equilibrium physics, will also serve as a starting
point in understanding the behavior of analogous systems in the
realm of nonequilibrium physics where the polarity of these
nematic phases is driven by active mechanisms.

K.S. acknowledges valuable discussion with John Toner.  W.K. and M.C.M.
were supported by NSF Grants DMR-0219292 and DMR-0305407.


\begin{references}
\bibitem{banana}
We should note that this is in contrast to molecules forming a smectic-C*
phase or in banana molecules, the dipole moment usually has a
component perpendicular to the molecular axis.
\end{references}
\end{document}